\documentclass [aps,prL,amssymb,amsmath,twocolumn,showpacs]{revtex4-1}
\usepackage{microtype}
\usepackage{amsmath}
\usepackage{amssymb}
\usepackage{amsthm}
\usepackage{mathrsfs}
\usepackage{graphicx}
\usepackage{verbatim}
\usepackage{bm}
\usepackage[dvipsnames]{xcolor}

\usepackage{graphicx,epsfig,amsfonts,amssymb}

\usepackage[colorlinks,
linkcolor=blue,
anchorcolor=blue,
urlcolor=blue,
citecolor=blue
]{hyperref}

\usepackage{bm}% bold math
\usepackage{times}
\usepackage{lipsum}

\usepackage[all]{xy}

\newcommand{\bk}{\mathbf{k}}

\newcommand{\bh}{\mathbf h}

\newcommand{\ve} {\varepsilon}

\newcommand{\be}{\begin{eqnarray}}
\newcommand{\ee}{\end{eqnarray}}
\newcommand{\la}{\langle}
\newcommand{\ra}{\rangle}
\newcommand{\rar}{\rightarrow}

\begin{document}

\title{Full investigation of nonadiabatic dynamical characterization in  arbitrary quenching process}

\author {Panpan Fang$^1$}
\author {Xinwei Shi$^1$}
\author{Yi-Xiang Wang $^{1,2}$} 
\email{wangyixiang@jiangnan.edu.cn}
\author{Fuxiang Li$^1$}
\email{fuxiangli@hnu.edu.cn}
\affiliation{$^1$School of Physics and Electronics, Hunan University, Changsha 410082, China}
\affiliation{$^2$School of Science, Jiangnan University, Wuxi 214122, China}

\date{\today}

\begin{abstract}
 Recently, dynamical characterization of bulk topology has been experimentally realized under nonadiabatic sudden quench dynamics.  However, it has been shown that only the topology of final phase can be characterized when the system is quenched from initial topologically trivial phase. In this paper, taking the two-dimensional Chern insulator as an example, we make a thorough investigation of different types of quenching processes under nonadiabatic slow quench dynamics, and study not only the processes between nontrivial phase and trivial phase, but also between the phases with different topological invariants.  We find that, under slow quench dynamics,  {\it both the initial and final topological phase} can be characterized and the topological invariant can be captured by time-averaged spin polarization.  Moreover, different types of processes can be distinguished from the special regions where the time-averaged spin polarization vanishes. All the dynamical characterization schemes are entirely based on the  experimentally measurable quantity time-averaged spin polarization, and thus one can expect our findings may provide reference for future experiments.
 \end{abstract}

%\pacs{05.60.-k, 05.40.-a, 82.37.-j, 82.20.-w}
%05.60.-k Transport processes
%05.40.-a Fluctuation phenomena, random processes, noise, and Brownian motion
%05.70.Ln nonequilibrium thermodynamics
%05.10.Gg Stochastic analysis methods (Fokker-Planck, Langevin, etc.)
%62.25.Fg High-frequency properties, responses to resonant or transient (time-dependent) fields
%82.20.-w Chemical kinetics and dynamics
%82.37.-j Single molecule kinetics

\date{\today}

\maketitle

%%%%%%%%%%%%%%%%%%%%%%%%%%%%%%%%%%%

%\begin{figure} [!htb]
%{\includegraphics[width=2\columnwidth]{fig1.png}}
%\caption{Adiabatic spectrum for BCS Hamiltonian (\ref{h-bcs0}) for different couplings $g=1/N_s$ (a) and $g=1$ (b). Total number of spins is $N_s=10$. The energy levels are $\ve_j=1-j/N_s$. } \label{fig:spectrum}
%\end{figure}
%
\section{Introduction}
Topological quantum phases have been  extensively studied in last two decades \cite{Hasan2010,qi2011,Rechtsman2013, Khanikaev, Ozawa, Miyake2013,Jotzu,W,S}. Under the equilibrium theory, the topological quantum phases can be classified and characterized by its topological invariant defined with Bloch functions. The famous bulk-boundary  correspondence dictates the existence of robust boundary mode that is immune to disorder or defects for a system with nontrivial topological invariant. According to the bulk-boundary correspondence, one can identify the topological phases by resolving the boundary modes with angle-resolved photoelectron spectroscopy and transport measurements experimentally \cite {Hsieh, Xia09, Chang13, Xu2015}.

Going beyond the equilibrium theory, the notion of characterization of topological phases has been extended to nonequilibrium regime, in which dynamical characterization of (non-) Hermitian quantum systems, (non-) correlated systems, higher-order  topological insulators are discussed \cite{2020winding, 2019correlate, 2004Floquet, 2021noise,2021highorder, Wang2017, Qiu2019, Xie2020, Hu2020, Chen2020, Song2019Nat}. Recently, a dynamical bulk-surface correspondence is established  by Liu and his coworkers in Hermitian systems \cite{2018Sci}, in which a generic $d$D topological phase can be characterized by the $(d-1)$ D invariant defined on the so-called  band inversion surface (BIS). On the platform of  optical lattice in ultracold atomic systems, the measurement of the time-averaged spin polarization (TASP) by spin-resolved time-of-flight absorption imaging becomes possible. Thus, this dynamical bulk-surface correspondence has been further verified experimentally \cite{experiment1, experiment2, experiment3,Liu2013, Liu2014}. In momentum space, after suddenly quenching the system from a trivial phase to a topological phase, the bulk topology of a $d$D equilibrium phase of the postquench Hamiltonian can be easily determined with high-precision by the winding of dynamical field on BIS in the TASP. In addition, a dynamical topological invariant after a sudden quench is proposed in Ref. \cite{Chen2018}, which is shown to be intrinsically related to the difference between the topological invariants  of the initial and final static Hamiltonian. The dynamical topological invariant is zero  if the initial Hamiltonian and the final Hamiltonian lie within the same phase. On the contrary, the dynamical topological invariant is not zero if the initial Hamiltonian and the final Hamiltonian lie in the different phases including different topological phases.

Compared with the above sudden quench, a general dynamical characterization scheme based on the slow quench protocol with a finite quenching rate is provided in  previous works \cite{2022us2,2020us1}. The quenching rate varies from $0$ to $\infty$, corresponding to a continuous crossover from the sudden quench limit to the adiabatic limit.  It has been found that, nonadiabatic slow quench dynamics can indeed provide an alternative topological structure named spin inversion surface (SIS) in characterizing the topological phases. However, the previous studies  consider only the dynamical characterization in one special type of quenching process,  and the TASP only captures the topological invariant of the postquench Hamiltonian. One may wonder whether it is critical for dynamical characterization to quench from a topologically trivial phase or whether the initial topological phase can be characterized. What knowledge could we obtain by reversing the quenching processes (i.e. from ``topological set of parameters'' to ``non-topological ones'') or by quenching between regimes with different topological invariants?   Therefore, by a  thorough investigation of the dynamical characterization for the two-dimensional Chern insulator, one would expect  richer physical phenomena emerging from different types of quenching processes.

In this paper, in the framework of slow nonadiabatic quench, we show how the dynamical characterization is performed in different types of quenching processes of the two-dimensional Chern insulator. A generalized version of two-level Landau-Zener model is proposed and applied to the study of dynamical characterization of topological phases under slow quench dynamics. By analyzing the corresponding TASP of each process after quenching, we find that the initial phase can be characterized under nonadiabatic quench dynamics. No matter what intermediate phase the quenching process undergoes, the TASP records the topological information of the initial phase and final phase. Specifically,  in the processes with a trivial phase included, the invariant of the initial or final topological phase can always be captured. In the processes between different topological regimes, both the initial and final phase can be characterized.  Finally, by analyzing the influence of the quenching rate and the ratio of relevant parameters, we find that the processes quenching from a topological phase to a trivial phase can always be distinguished from other type of processes. Compared with the sudden quench, this is also the unique advantage of slow quench. All the dynamical characterization schemes are entirely based on the TASP, which can be observed directly in experiment. Therefore, one can expect our findings may provide reference for future experiments.

The rest of this paper is organized as follows. In Sec.~\ref{section:2i}, we give an general  introduction to the slow nonadiabatic quench dynamics. Then in Sec.~\ref{section:3i}, we present the results of dynamical characterization in different types of quenching processes under nonadiabatic slow quench dynamics.  In addition, we discuss the influence of the quenching rate and the ratio of interactions to the TASP in Sec.~\ref{section:3i}. Finally, we provide a brief summary to the main results of the paper in Sec.~\ref{section:4i}.

\section{Nonadiabatic quench dynamics}\label{section:2i}
\subsection{Nonadiabatic slow quench dynamics}
We first give a general introduction to the slow quench protocol that can be utilized in the nonadiabatic dynamical characterization of topological phases. In general, for slow quench, the Hamiltonian of a topological system in momentum space is described in the following form
\be
{\cal H }(\bm k,t) = {\bh}(\bm k,t)\cdot {\bm \gamma} =h_{0}(\bm k,t) \gamma_0+ \sum_{i=1}^{d} h_{i}(\bm k) \gamma_i. \label{H1}
\ee
The ${\bm \gamma}$ matrices  satisfy the anticommutation relations $\{\gamma_i, \gamma_j\}=2\delta_{ij}$ and are of dimensionality $n_d = 2 ^{d/2}$ (or $2^{(d+1)/2}$) when $d$ is even (or odd). In 1D and 2D systems, ${\bm \gamma}$ are the Pauli matrices. In higher dimensional systems,  ${\bm \gamma}$  take the Dirac form.

Here, we consider a specific protocol:
 $$h_0(\bm k,t) = g/t + h_0(\bm k)$$
  with $g$ determining the  quenching rate. The parameter $g$ varies from $0$ to $\infty$, corresponding to a continuous crossover from the sudden quench limit ($g=0$) to the adiabatic limit ($g\rar \infty$). In such a protocol, the form of $g/t$ enables us to quench the system from an initial Hamiltonian at $t=t_{int}$ to a final Hamiltonian at $t=t_f$ by keeping other parameters unchanged. During the quenching process, the evolution of the state vector is fully determined by the Schr\"odinger equation  $i\hbar\frac{d}{dt}|\psi(\bm k,t)\ra={\cal{H}}({\bm k},t)|\psi(\bm k,t)\ra$ after preparing an initial state $|\psi(\bm k,t_{int})\ra$. Then the pseudospin defined as
\be
{\la {\bm \gamma}(\bm k,t) \ra}=\la \psi(\bm k,t) | {\bm \gamma}| \psi(\bm k,t)\ra
\ee 
 will precess about the effective field ${\bh}(\bm k,t)$. After a certain time $\Delta t$ [$g/(t_{int}+{\Delta} t)\ll h_0(\bm k)$],
 one can observe an approximately stable oscillation of the pseudospin ${\la {\bm \gamma}(\bm k,t) \ra}$. 
 Thus, the final TASP over a period $T$ at each $\bm k$ point can be obtained as:
\be
 \overline{\la {\bm \gamma}(\bm k) \ra} = \frac{1}{T}\int_{t_{int}+{\Delta} t}^{t_{int}+{\Delta} t+T}{\la {\bm \gamma}(\bm k,t) \ra} dt.
 \ee
Here, the time point $t_{int}+{\Delta} t$ across the time point of phase boundary at ${\bh}(\bm k,t)=0$. The superscripts/subscripts $``f"$ and $``int"$ here represent the parameters of the final (postquench) Hamiltonian ${\cal H}_f$ and initial (prequench) Hamiltonian  ${\cal H}_{int}$, respectively. 
\subsection{The TASP under slow quench dynamics}
To be more clear about the expression form of TASP, one needs to consider the nonadiabatic dynamics governed by the time-dependent Hamiltonian (\ref{H1}), for which purpose we need to solve the corresponding Landau-Zener problem.  Specifically, for a two-level system, if one prepares an initial state $|\psi(\bm k,t_{int})\ra$ as the ground state of initial Hamiltonian ${\cal H}_{int}$, the system  will undergo a nonadiabatic transition during the evolution, and finally at time $t= t_f$, the system will  stay not only on the final instantaneous ground state $|-\ra $ with  probability $P_d(\bm k)$, but also on the instantaneous excited state $|+\ra$ with probability $P_u(\bm k)$. 
By solving the time-dependent Schr\"odinger equation, one can find that each component of TASP has the following form (see Appendix.~\ref{appendix A}):
\be
\overline{\la {\gamma_i}(\bm k) \ra} = (P_u-P_d) \frac{h_i^f}{\ve_f}
\ee
in which $\ve_f$ is the eigenenergy of the final Hamiltonian and $h_i^f$ is the component of effective vector field ${\bh}(\bm k,t)$ in Eq.~(\ref{H1}) at $t=t_f$. Note that the transition probability $P_u$ and $P_d$, parameters $h_i^f$ and $\ve_f$ are all dependent on momentum $\bm k$. 

One would find that each component of TASP can vanish in two special regions in the $\bm k$ space: one is $h_i=0$, the other is $(P_u-P_d)=0$. For the region with $h_i=0$, the component $\overline{\la {\gamma_i}(\bm k) \ra}$ of TASP vanishes but other components may remain nonzero and  we define it as BIS. For the region with  $P_u-P_d=0$, all components of TASP vanish and we define it as SIS. Specifically, if we choose $h_0$ as the quenching axis, and thus in the $\overline{{\la { \gamma}_0 (\bm k) \ra}}$, the region with $h_0^f=0$ corresponds to the BIS and the region with $P_u-P_d=0$ corresponds to the SIS. 

Physically, the appearance of $P_u-P_d=0$ may be caused by the properties of either initial or final state. However, the transition probability $P_u$ and $P_d$ may sometimes be sensitive to both initial and final state of the nonadiabatic transition, and thus SIS here may be subdivided into two parts: one is FSIS, with its position close to the BIS; the other is ISIS, with its position far away from the BIS. Both the BIS and SIS will be identified after experimentally obtaining  the measurable TASP. The information related to the BIS and SIS is always marked in purple and green in the following description, respectively.

%---------------------------------------------------------
\section{different types of quenching processes for the  2D topological insulator }\label{section:3i}
Now we apply different types of quenching processes to the 2D Chern insulator, which is generically described by a two-band Hamiltonian: ${\cal H}(\bm k,t) = \bh(\bm k,t) \cdot {\bm \sigma}$, with the vector field given by:
\be
&&h_0(\bm k,t)\equiv{h_z(\bm k,t)} = \frac{g}{t}+m_z - t_0 \cos k_x - t_0 \cos k_y,\nonumber\\
&&h_{1}=h_x=t_{so} \sin k_x, \label{eqh1}\\
&&h_{2}=h_y=t_{so} \sin k_y\nonumber.  
\ee

This Hamiltonian without time-dependent term $\frac{g}{t}$ has been realized in recent experiment of quantum anomalous Hall effect \cite{QAH}. Here, we slowly quench the $z$-component of vector field from $t=t_{int}$ to $t_f$ (a large number compared with $g$), and study the emergent topological characterization after quenching.  For $0<\frac{g}{t}+m_{z}<2t_0$, the  Hamiltonian gives a topological phase with Chern number ${\cal C}=-1$. In addition, for $-2t_0 <\frac{g}{t}+ m_{z}<0$, the Hamiltonian describes a topologically nontrivial phase with Chern number ${\cal C}=+1$. Otherwise, the Hamiltonian lies in the trivial phase. 

In the following, we will first discuss the TASP in different types of quenching processes under slow quench dynamics with a certain $g$ and a certain ratio $t_{so}/t_0$, and then discuss the influence of the quenching rate $g$ and the ratio $t_{so}/t_0$ on the TASP.

\subsection{quenching the system from a trivial phase to a topological phase}
%---------------------------------------------------------------------------------------------------
\begin{figure}[htbp]
	\centering
	\epsfig{file=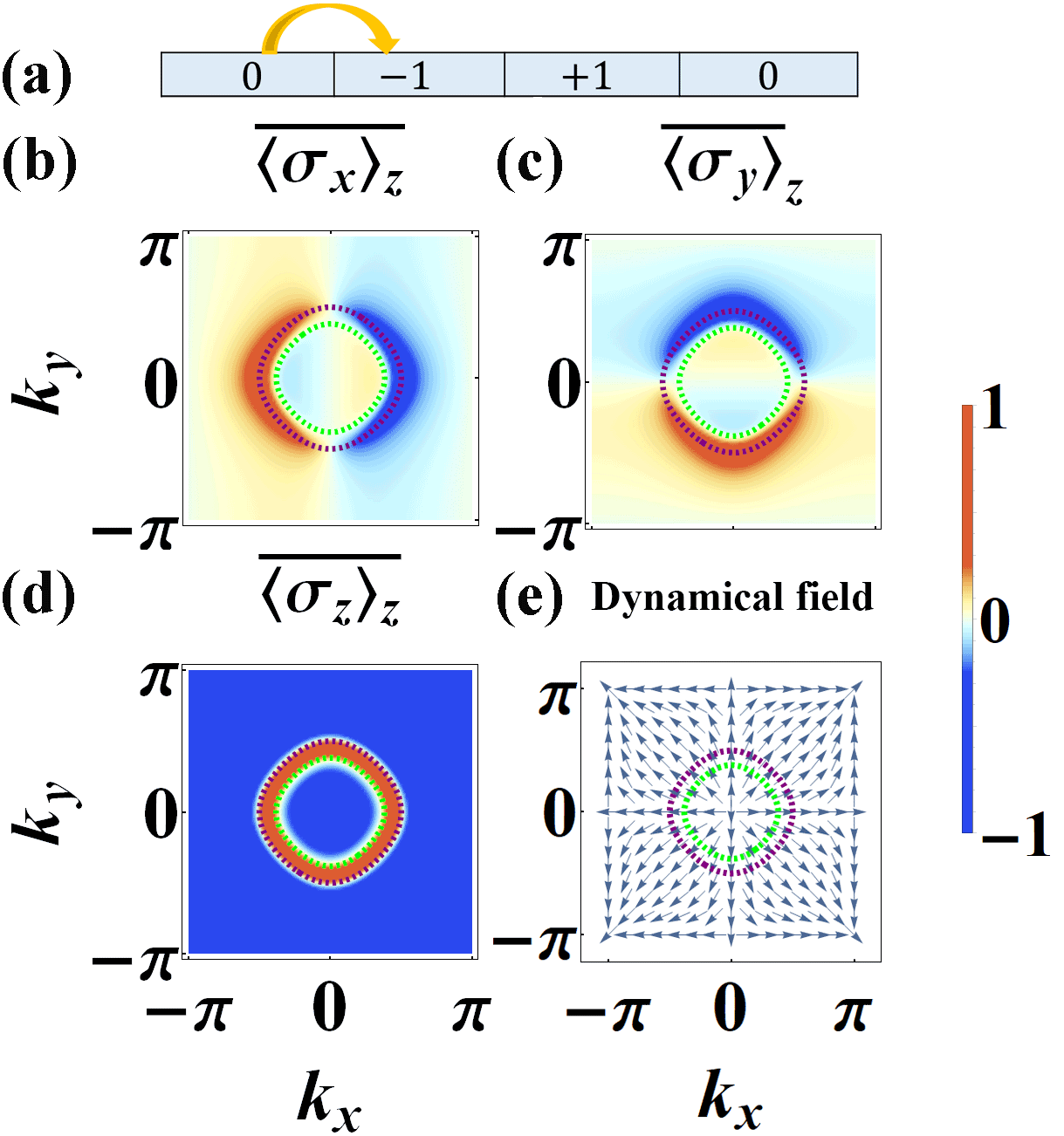, width=3.4in}
	\caption{(a) The specific quenching processes of slow quench. "0" represents the trivial phase, and other numbers represent the topological phase with different topological invariant. (b)-(d) The TASP after quenching $h_z$ axis from a trivial phase to a topological phase with $C=-1$. The system is quenched from $t_{int}=0$ to $t_f=5000$ with $5t_{so}=t_0=g=m_z=1$.  $\overline{\la { {\sigma}_z}\ra_z}$ represents the $z$ component of TASP after quenching $h_z$ axis, and so on. (e) The dynamical field of the system plotted by the normalized spin-orbit field $\hat{\bm h}_{so}$.}
	\label{fig:tr_t1}
\end{figure}
\begin{figure}[htbp]
	\centering
	\epsfig{file=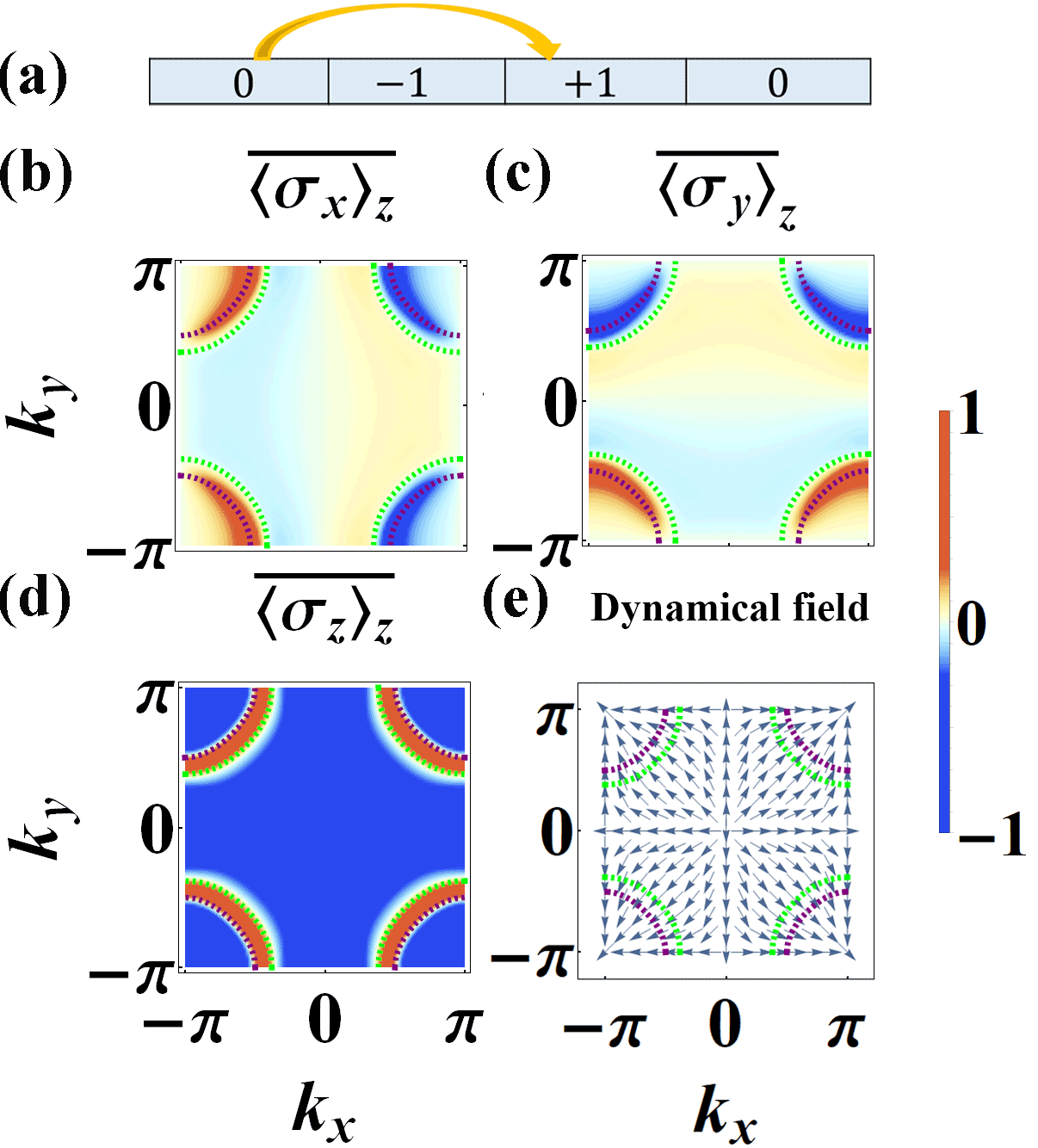, width=3.4in}
	\caption{(a) The specific quenching processes of slow quench. "0" represents the trivial phase, and other numbers represent the topological phases with different topological invariants. (b)-(d) The TASP after quenching $h_z$ axis from a trivial phase to a topological phase with $C=+1$. The system is quenched from $t_{int}=0$ to $t_f=5000$ with $5t_{so}=t_0=g=-m_z=1$. (e) The dynamical field of the system plotted by the normalized spin-orbit field.}
	\label{fig:tr_t2}
\end{figure}
 The exact solutions of the Landau-Zener problem are difficult to find and only a few special cases can be solved exactly.
In the previous paper \cite{2022us2}, we have given an exact solutions in a special case. Briefly speaking, we slowly quench the system from a trivial phase to a topological phase with time $t$ varies from $t_{int}\rar0$ to $t_f\rar\infty$, and the initial state
$|\psi(0)\ra$ is always the ground state of the initial Hamiltonian ${\cal {H}} (t_{int}\rar0)$. Then the TASP can be given by 
\be
\overline{\la { {\sigma}_i}\ra} = (P_u-P_d)\frac{{h_i^f}} {\varepsilon_f}=\frac{({e^{-2\pi g \frac{h_0}{\ve_f}}-\cosh{2\pi g}})}{\sinh{2\pi g}}\frac{{h_i^f}} {\varepsilon_f}.\quad \label{sigma1}
\ee
 with ${\ve}$ is the energy of final Hamiltonian $H(t_{int}\rar\infty)$.

  As shown in Figs.~\ref{fig:tr_t1}(b), ~\ref{fig:tr_t1}(c), and ~\ref{fig:tr_t1}(d),  we plot the three components of TASP after quenching the $h_z$ axis from a trivial phase to a topological phase with topological invariant $-1$.
  There are two rings with $\overline{\la { {\sigma}_z}\ra}=0$ appearing in the TASP. On the purple ring,  only one component of TASP $\overline{\la { {\sigma}_z}\ra}$ vanishes, while the other two components remain nonzero. 
   On the green ring, all three components of TASP vanish, which corresponds to the momentum points with $P_u-P_d=0$ in Eq.~(\ref{sigma1}). Thus, we identified the purple ring and green ring as BIS and SIS, respectively. After identifying the BIS and SIS in the TASP, one can then determine the topological invariant of system from the winding of dynamical field formed by the other two components $-\overline{\la { {\sigma}_{x}}\ra}$ and $-\overline{\la { {\sigma}_{y}}\ra}$ on BIS, or by the gradients of two other components ${\widetilde{{g_{x}}}}=-_{\partial{k_{\perp}}}{\overline{\la \sigma_{x}\ra}}$ and ${\widetilde{{g_{y}}}}=-_{\partial{k_{\perp}}}{\overline{\la \sigma_{y}\ra}}$ on SIS \cite{2020us1}. Here,  the direction $k_{\perp}$ is defined to be perpendicular to the SIS. Both the dynamical field $(-\overline{\la { {\sigma}_{x}}\ra}, -\overline{\la { {\sigma}_{y}}\ra})$ and $({\widetilde{{g_{x}}}},{\widetilde{{g_{y}}}})$ are shown to be proportional to the spin-orbit field $\bm {h_{so}}=(h_x,h_y)$. That is to say, $\bm {h_{so}}$ is actually a two components vector, as shown in Fig.~\ref{fig:tr_t1}(e). One can see that both the winding of dynamical field (arrows) on BIS and SIS forms a nontrivial topological configuration, and thus gives a topological number $-1$. 
  
  In Figs.~\ref{fig:tr_t2}(b), ~\ref{fig:tr_t2}(c), and ~\ref{fig:tr_t2}(d), three components of TASP are shown after quenching the $h_z$ axis from a trivial phase to a topological phase with topological invariant $+1$. The purple ring and the green ring can also be identified as BIS and SIS, respectively. Furthermore, one can also obtain the dynamical field on BIS and SIS by measuring the value and the gradient of two components of TASP $\overline{\la { {\sigma}_{x,y}}\ra}$ , respectively. Compared with the dynamical field shown in Fig.~\ref{fig:tr_t1}(e), the dynamical field plotted in Fig.~\ref{fig:tr_t2}(e), which converges at a central point inside BIS and SIS rather than spread out to the outside of BIS and SIS, shows an opposite winding behavior.  Therefore, an opposite Chern number  $C=+1$ is given by the dynamical field both on BIS and SIS.

 From the above results of Fig.~\ref{fig:tr_t1} and Fig.~\ref{fig:tr_t2}, one can see that what  the TASP  records is the topological invariant of the final phase. Even though in the case of Fig.~\ref{fig:tr_t2}, the system undergoes a nonadiabatic transition that passes through an intermediate region with Chern number $C=-1$, the information of intermediate phase is unrevealed.
 In next subsections, we will show that the topological invariant of initial phase can also be revealed in the TASP. However, in the current case, the initial trivial phase  cannot bring any topological information on the TASP. Thus, only the information related to final topological phase is recorded in the TASP.

\subsection{quenching the system from a topological phase to a trivial phase}
 We further explore the dynamical characterization in other types of quenching processes, in which the system is quenched from a topological phase with topological invariant $C= +1$ to a trivial phase.  As shown in Fig.~\ref{fig:t_t}(b), there is only one ring ( denote by green solid line),  on which all the components of TASP vanish, and thus we also identify this ring as SIS. On the SIS, we can still obtain the dynamical field as shown in Fig.~\ref{fig:tr_t2}(e), which implies a topological number $ +1$. Evidently, the final trivial phase here cannot bring any topological information on the TASP. Thus, the TASP here only record the topological information related to the initial phase. Note that, here we use the green solid line  in order to show that the appearance of SIS is due to a topological initial phase. If the appearance of SIS is due to a topological final phase, then we will still denote the SIS by green dashed line. 
 
\subsection{quenching the system from a topological phase to a topological phase}{\label{section:3ii}}
%---------------------------------------------------------------------------------------------------
\begin{figure}[htbp]
	\centering
	\epsfig{file=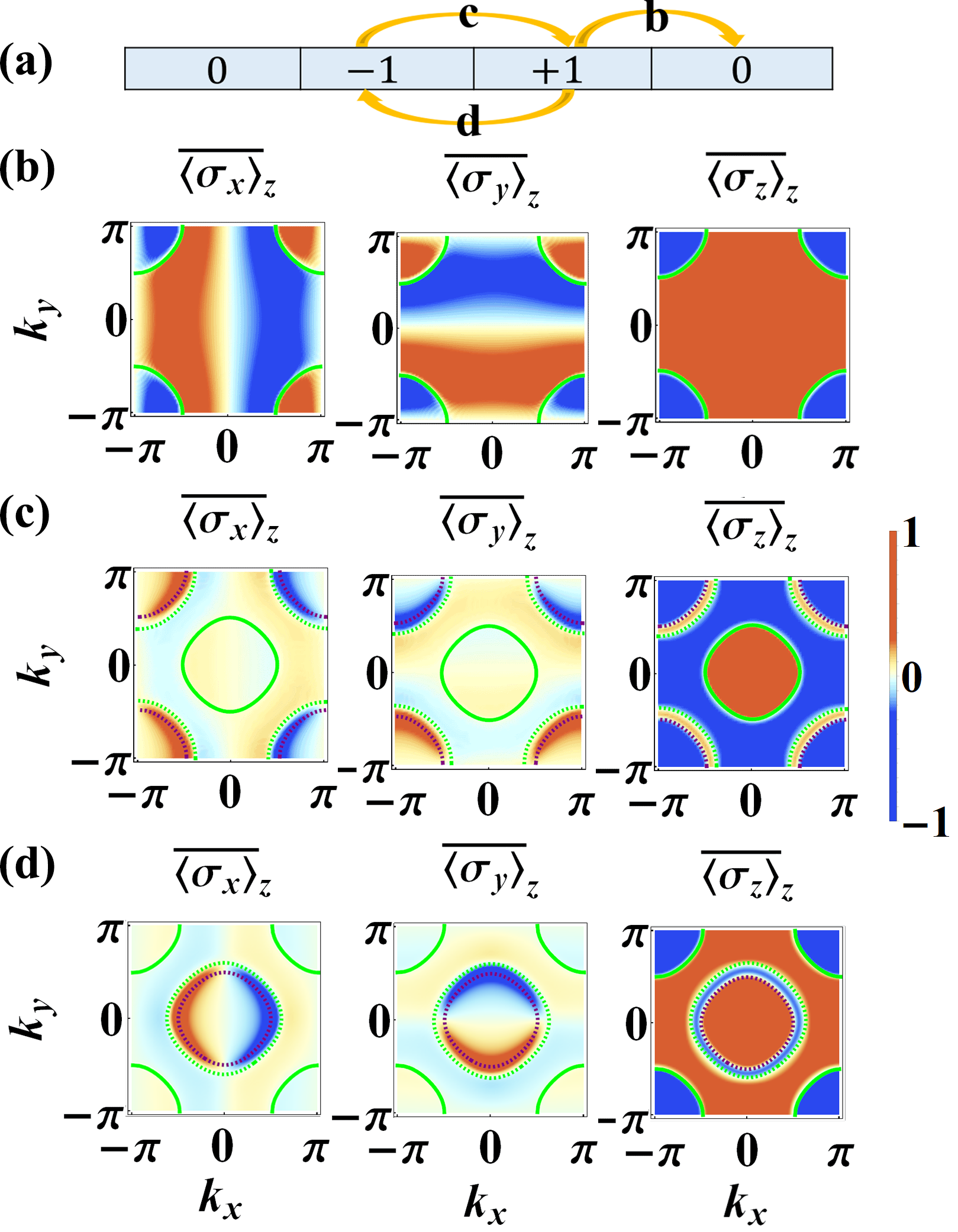, width=3.4in}
	\caption{(a) The specific quenching processes in (b), (c) and (d). "0" represents the trivial phase, and other numbers represent the topological phases with different topological invariants. (b) The TASP after quenching $h_z$ axis from a topological phase with $C=+1$ to a trivial phase. The system is quenched from $t_{int}=1/3$ to $t_f=5000$ with $5t_{so}=t_0=g=1$ and $m_z=-4$. (c) The TASP after quenching $h_z$ axis from a topological phase with $C=-1$ to another topological phase with $C=+1$. The system is quenched from $t_{int}=1/2$ to $t_f=5000$ with $5t_{so}=t_0=g=-m_z=1$.   (d) The TASP after quenching $h_z$ axis from a topological phase with $C=+1$ to another topological phase with $C=-1$. The system is quenched from $t_{int}=1/2$ to $t_f=5000$ with $5t_{so}=t_0=-g=m_z=1$.}
	\label{fig:t_t}
\end{figure}

%---------------------------------------------------------------------------------------------------
 Now, we discuss a different type of quenching process in which the system is quenched from a topological nontrivial phase to another nontrivial phase. 
 
 We first quench the system from a topological phase with topological invariant $C=-1$ to another topological phase with topological invariant $C=+1$. As shown in Fig.~\ref{fig:t_t}(c), SIS here is obviously subdivided into two parts, and thus there appear three different rings with $\overline{\la { {\sigma}_z}\ra}=0$ in the TASP. To distinguish them, we name this two parts as FSIS and ISIS.  Near the edge of Brillouin zone, there are two rings (green dashed and purple dashed) adjacent to each other, which are identified as BIS and FSIS. They can be further distinguished by different behaviors of  two other components of TASP $\overline{\la { {\sigma}_{x,y}}\ra}$. On the BIS, as denoted by purple dashed ring, the other two components of TASP $\overline{\la { {\sigma}_{x,y}}\ra}$ remain nonzero. On the FSIS, as denoted by green dashed ring, the other two components of TASP $\overline{\la { {\sigma}_{x,y}}\ra}$ also vanish. The third ring, which is away from BIS and near the edge of Brillouin zone, is identified as ISIS on which all the components of TASP also vanish. Then one can also obtain the dynamical field on the BIS or FSIS by measuring the value or the gradient  of two components of TASP $\overline{\la { {\sigma}_{x,y}}\ra}$ . Both the dynamical field on BIS and FSIS have same configuration as shown in the Fig.~\ref{fig:tr_t2}(e). Thus, a nontrivial topological number $+1$, which corresponds to the topological invariant of the final phase, is  given by the dynamical field on BIS or FSIS. Moreover, one can also obtain the dynamical field on ISIS by measuring the gradient of two components of TASP $\overline{\la { {\sigma}_{x,y}}\ra}$. The dynamical field on ISIS have similar topological configuration as  shown in Fig.~\ref{fig:tr_t1}(e). Thus, a nontrivial topological number $-1$, which corresponds to the topological invariant of the initial phase, is given by the dynamical field on the ISIS. 
 
 The above characterization schemes remain valid for the quenching process from another topological phase with topological invariant $C=+1$ to a topological phase with topological invariant $C=-1$. In order to reverse the quenching process, we change the quench protocol $g/t$ as $-g/t$.  As shown in Fig.~\ref{fig:t_t}(d), the two adjacent rings  also appear in the TASP. Similar to the quenching processes as before, we can identify one ring as the BIS and the other as FSIS. On both of them, the dynamical field presents the same topological configuration as  shown in Fig.~\ref{fig:tr_t1}(e), which implies the bulk topological number of the final phase $C=-1$. In addition, we can still identify the remaining ring as ISIS, on which all the  components of TASP vanish. The dynamical field  on it presents the same topological configuration as that shown in Fig.~\ref{fig:tr_t2}(e), which implies the bulk topological number of the final phase $C=+1$.

From the perspective of phase transition, in the process of slow quench,  $h_z(\bk,t)$ varies with time, and the bulk gap closes while $h_z(\bk,t) = { {h_{x}}}=h_y=0$,  corresponding to a topological phase transition with the topological invariant changed. In particular, ${{h_{x}}}=h_y=0$ are the positions of topological charges, i.e., the singularities enclosed in the ISIS, FSIS, and BIS. Analogous to the Gaussian theorem \cite{2019charge, 2021highcharge}, the topological charge is actually a monopole, whose quantized flux through the ISIS and FSIS$/$BIS are viewed as the initial topological invariant and final topological invariant, respectively. That is to say, the topological charge is dual to the winding of dynamical field on the ISIS, FSIS and BIS. Thus, one can also determine the initial and final topological invariants of the system from the corresponding topological charges. (see Appendix.~\ref{appendix B} for detail).

Up to now, we find that the initial phase can also be characterized under nonadiabatic quench dynamics. No matter what intermediate phase the quenching process undergoes, the TASP  only records  the topological information of the initial phase and final phase. In addition,  each type of quenching process shows its unique features on the SIS and BIS, and thus the different types of processes can be distinguished.  In particular, the initial phase and final phase can be easily characterized and distinguished in all types of quenching processes. Similar results can also be obtained in the topological system with high Chern number (see Appendix.~\ref{appendix C}).

\section{the influence of the quenching rate $g$ and the ratio $t_{so}/t_0$}\label{section:4i}
In this section, we will  give a more general conclusion after considering the influence of the quenching rate $g$ and the ratio $t_{so}/t_0$ on the dynamical characterization.

\subsection{the influence of the quenching rate $g$}\label{section:4i1}

\begin{figure}[htbp] 
	\centering
	\epsfig{file=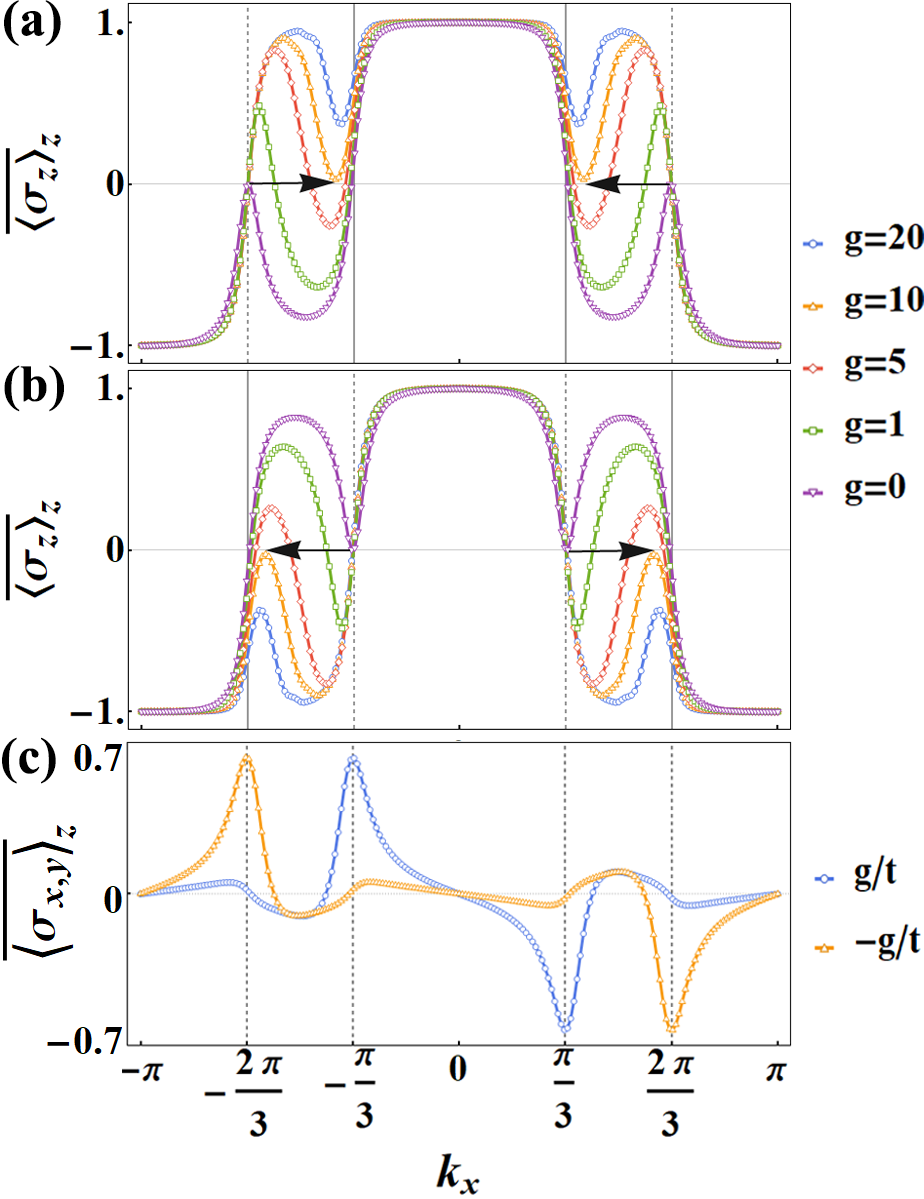, width=3.4in}
	\caption{(a) The component of TASP $\overline{\la {{\sigma_z}}\ra}$ on the line $k_x=k_y$ in the processes between different topological regimes with quench protocol $g/t$. The system is quenched from a topological phase with $C=-1$ to a topological phase with $C=+1$. Other parameters satisfy $(g/t_{int}+m_z)=1$, $(g/t_{f}+m_z)=-1$ and $5t_{so}=t_0=-m_z=1$.  (b) The component of TASP $\overline{\la {{\sigma_z}}\ra}$ on the line $k_x=k_y$ in the processes between different topological regimes with quench protocol $-g/t$. The system is quenched from a topological phase  with $C=+1$ to a topological phase with $C=-1$.  Other parameters satisfy $(g/t_{int}+m_z)=-1$, $(g/t_{f}+m_z)=+1$ and $5t_{so}=t_0=m_z=1$. The solid and dashed lines represent the points with $h_z^{int}=0$ and $h_z^f=0$ in corresponding process, respectively. The change of FSIS are denoted by the arrows. (c) The component of TASP $\overline{\la {{\sigma_{x,y}}}\ra}$ on the line $k_x=k_y$ in the processes between different topological regimes with two quench protocol $g/t$ and $-g/t$ ($g=1$).}
	\label{fig:gg2}
\end{figure}
As shown in Fig.~\ref{fig:gg2} (a) and (b), for the processes between different topological regimes, we plot the component of TASP $\overline{\la {{\sigma_z}}\ra}$ as a function of quenching rate $g$ on the line $k_x=k_y$  with protocol $g/t$ and $-g/t$, respectively. Thus, the BIS, FSIS, and ISIS can be identified as before under a certain rate $g$. The solid and dashed lines represent the zero points of  $h_z^{int}$ and $h_z^f$ in corresponding process, respectively. 

When $g$ changes, the BIS is always at momentum points with $h_z^f=0$ as defined before. However, the position of SIS depends on $g$. Although the position of ISIS changes more slowly with $g$ than the FSIS, they move in opposite directions (face to face) as $g$ becomes larger. Here, we denote the change of the position of FSIS by the arrows. Thus, when $g$ is large enough ($g=10$), the FSIS will overlap with ISIS. In addition, the exact form of TASP in sudden quench is given as (see Appendix.~\ref{appendix A}):
\be
\overline{\la {{\sigma_i}}\ra}=-\frac{(h_0^fh_0^{int}+h_1^2+h_2^2)h_i^f}{\ve_{int}\cdot\ve_{f}^2}.
\label{eq:sq1}
\ee
in which $h_0,h_1$, and $h_2$ equal to $h_z,h_x$, and $h_y$, respectively. 

When $g$ tends to zero, the nonadabatic slow quench here degenerates into  sudden quench. Thus, the position of FSIS and ISIS will totally overlap with two zero points of $h_0^fh_0^{int}+h_1^2+h_2^2$. The smaller the $(h_x^2+h_y^2)$ is, the closer the FSIS and ISIS is to $h_z^{f}=0$ and $h_z^{int}=0$. In a certain range $0\leq g<10$, there will be a FSIS near the BIS and a ISIS far from the BIS, on which the dynamical field reflects the bulk topology of the final phase and initial phase, respectively.

For the processes with quench protocol $-g/t$, it still can be considered as a Landau-Zener problem. Thus, the TASP of $-g/t$ also has the similar form $\overline{\la {\sigma_i}(\bm k) \ra} = (P_u-P_d) \frac{h_i^f}{\ve_f}$ as that of $g/t$. For the case here, after further plotting the other components of TASP $\overline{\la {{\sigma_{x,y}}}\ra}$ in Fig.~\ref{fig:gg2} (c), one can see that the TASP of $g/t$ and $-g/t$ have the following relation :
\be
\overline{\la { {\sigma}_i(\bm k)}\ra}_{g/t}= -\overline{\la { {\sigma}_i(\bm k+\pi)}\ra}_{{-g/t}}. 
\ee
 To make it more clear, we give a  schematic diagram of translation between this two processes in Appendix.~\ref{appendix B}. Thus,  one can easily obtain $[P_u(\bm k)-P_d(\bm k)]_{g/t}$=$[P_u(\bm k+\pi)-P_d(\bm k+\pi)]_{-g/t}$. Especially, the above relation can be well understood in sudden quench according to Eq.~\ref{eq:sq1}. For other cases, there is not an obvious general relation that can be found in the TASP and $(P_u-P_d)$ between two reversed processes, but the position of SIS depends on the $g$ is similar to the case described here. 
 
 In addition, the dependence of position of SIS on the $g$ in the processes from a trivial phase to a topological phase and from a topological phase to a trivial phase is similar to  the FSIS and ISIS, respectively, and we will not bother to elaborate on them.
%---------------------------------
\subsection{the influence of the ratio $t_{so}/t_0$}\label{section:4i2}
\begin{figure}[htbp]
	\centering
	\epsfig{file=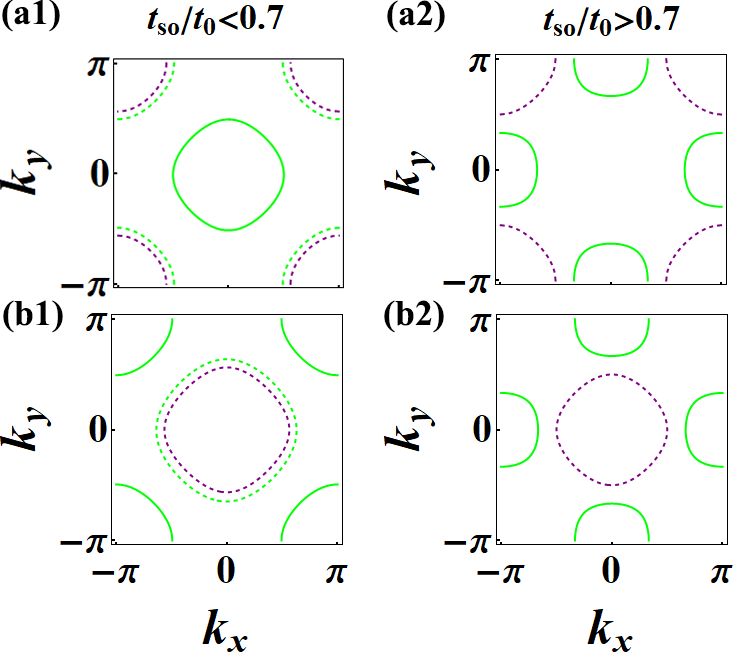, width=3.4in}
	\caption{(a1-a2)  After quenching the system from a topological phase with $C=-1$ to another topological phase with $C=+1$, two configurations of rings with different ratio $t_{so}/t_0$ are found in $\overline{\la {{\sigma_z}}\ra}$ for protocol $g/t$ $(g=1)$. (b1-b2) After quenching the system from a topological phase with $C=+1$ to another topological phase with $C=-1$, two configurations of rings with different ratio $t_{so}/t_0$ are found in $\overline{\la {{\sigma_z}}\ra}$ for protocol $-g/t$ $(g=1)$. }
	\label{fig:tso}
\end{figure}

In the quenching Hamiltonian ~(\ref{eqh1}), there is an additional parameter $t_{so}$ which does not affect the phase diagram but may influence the results of nonadiabatic transitions. In this subsection, we will study the effect of ratio $t_{so}/t_0$ on the dynamical characterization.
As plotted in Fig.~\ref{fig:tso}(a) and Fig.~\ref{fig:tso}(b), after quenching the system from a topological phase to another topological phase with protocol $g/t$ and $-g/t$, respectively, two configurations of rings with different ratio $t_{so}/t_0$ are found in $\overline{\la {{\sigma_z}}\ra}$, respectively.

 For the case of $t_{so}/t_0>0.7$, one can see that the SIS here decomposes into four semicircles. Similar to the ring in the center of TASP with $t_{so}/t_0<0.7$, two semicircles (up and down, left and right) here compose a ring, which gives a  invariant $-1$. In addition, when $g$ changes, the configuration of SIS remains unchanged in the two reversed quenching processes. Thus, either the initial or final invariant of system cannot be exactly characterized by SIS.  However, the components of TASP $\overline{\la {{\sigma_x}}\ra}$ and $\overline{\la {{\sigma_y}}\ra}$ on BIS remain nonzero.  Thus, the final phase can still be characterized.  

 To understand the underlining physics of different configurations of rings, one should study the vanishing TASP under slow quench dynamics with  small $g$, which  can be approximately obtained by the regions with momentum points determined by $h_0^fh_0^{int}+h_1^2+h_2^2=0$ from sudden quench dynamics.
 For convenience, we denote the regions with momentum points $h_0^fh_0^{int}+h_1^2+h_2^2=0$ as SIS$_{g=0}$. When $t_{so}/t_0$ approaches zero, the SIS$_{g=0}$ is actually in the region with momentum points $h_0^f=0$ and $h_0^{int}=0$. When $t_{so}/t_0$ becomes very large, the SIS$_{g=0}$ is no longer in the region with momentum points $h_0^f=0$ and $h_0^{int}=0$. Thus, there must be a critical value of $t_{so}/t_0$ at which the position of SIS changes abruptly. Under this critical value, the position of SIS$_{g=0}$ is close to the region with $h_0^f=0$ and $h_0^{int}=0$ and the shape of SIS$_{g=0}$ is also similar to the region with $h_0^f=0$ and $h_0^{int}=0$. The magnitude of this critical value is proportional to $m_z$ (See Appendix.~\ref{appendix A} for detail). Moreover, in the two reversed quenching processes, the product of the $h_0^f$ and $h_0^{int}$ keeps unchanged, and thus the same configuration of SIS$_{g=0}$ can be found in the TASP.

We further discuss the influence of $t_{so}/t_0$ in other types of quenching processes with a trivial phase included. In the quenching  processes from a trivial phase to a topological phase, the vanishing TASP in $\overline{\la {{\sigma_z}}\ra}$ can be found at SIS$_{g=0}$ and at BIS with $h_z^f=0$. Thus, there will be a special case that the SIS$_{g=0}$ sometimes may overlap with BIS (see Discussion~\ref{section:discussion}). Nevertheless, the BIS with $h_0^f=0$ always exits, and thus the final bulk topology in this processes can always be captured. 

\begin{figure}[htbp]
 	\centering
 	\epsfig{file=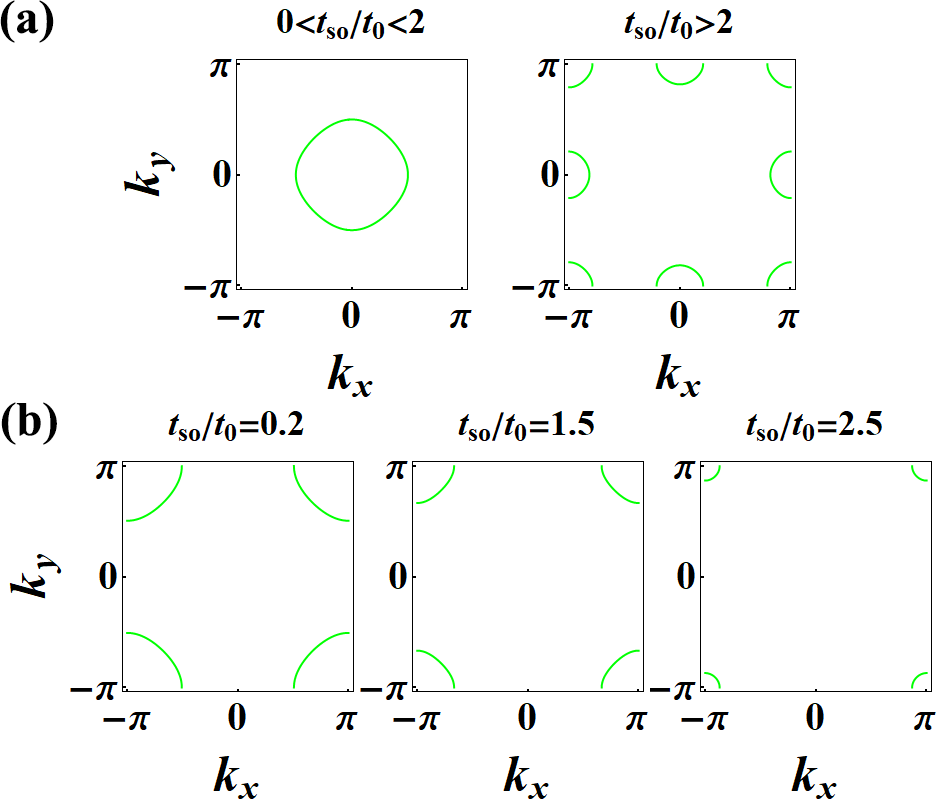, width=3.4in}
 	\caption{(a) The component of TASP $\overline{\la {{\sigma_z}}\ra}$ after quenching $h_z$ axis from a topological phase with $C=-1$ to a trivial phase for different ratio $t_{so}/t_0$. The system is quenched from $m_z=+1$  to $m_z=-3$. (b) The component of TASP $\overline{\la {{\sigma_z}}\ra}$ after quenching $h_z$ axis from a topological phase with $C=+1$ to a trivial phase for different ratio $t_{so}/t_0$. The system is quenched from $m_z=-1$  to $m_z=-3$.}
 	\label{fig:sqf}
 \end{figure}

In the quenching processes from a topological phase to trivial phase, $h_0^{f}$ cannot equal to zero, and thus the vanishing TASP in $\overline{\la {{\sigma_z}}\ra}$ can only be found at SIS$_{g=0}$.
As shown in Fig.~\ref{fig:sqf}, we plot the component of TASP $\overline{\la {{\sigma_z}}\ra}$ after quenching $h_z$ axis from a topological phase with $C=-1$ or $C=+1$  to a trivial phase for different ratio $t_{so}/t_0$. Only one configuration of SIS$_{g=0}$ is found in the processes from a topological phase with $C=+1$ to a trivial phase, and we have already known the invariant defined on it is $+1$. Two configurations of SIS$_{g=0}$ are still found in the processes from a topological phase with $C=-1$ to a trivial phase. However, because the $h_0^{f}$ is different from that in the processes between different topological regimes, the critical ratio is different. The first SIS$_{g=0}$ is an enclosed rings, which gives the initial topological number $-1$. For the second SIS$_{g=0}$, the semicircles at four corners of the Brillouin zone compose a ring that gives a topological number $+1$. The two pairs of semicircles (up and down, left and right) compose two rings with invariant $-1$, and one of them cancel out with the ring composed by four semicircles at four corners. Thus, the initial topological number $-1$ can always be obtained.  

Overall, the initial phase can also be characterized under nonadiabatic quench dynamics. No matter what intermediate phase the quenching process undergoes, the TASP records the topological information of the initial phase and final phase. Specifically,  in the processes with a trivial phase included, the invariant of the initial topological phase or final topological phase can always be captured. However, in the processes between different topological regimes, both the initial and final phase can be characterized under a critical ratio $t_{so}/t_0$. Beyond this critical ratio, only the final phase can be characterized. 

 Our findings also show that, under the critical value, each type of quenching process shows its unique features on the SIS and BIS.   Specifically, if only two adjacent rings appear in the TASP, we identify this process as quenching from a trivial phase to a topological phase.  If only one ring appears in the TASP, we identify this process as quenching from a topological phase to a trivial phase.   In addition to the above two cases, other types of processes are identified as quenching between different topological phases. Beyond this critical ratio (here, this critical ratio is 0.7), the configuration of rings is complicated. However, because only SIS appears in the processes from a topological phase to a trivial phase, this type of processes can always be distinguished from other types of processes in spite of what value the $t_{so}/t_0$ is.

%---------------------------------
\section{Discussion}\label{section:discussion}
\begin{figure}[htbp]
	\centering
	\epsfig{file=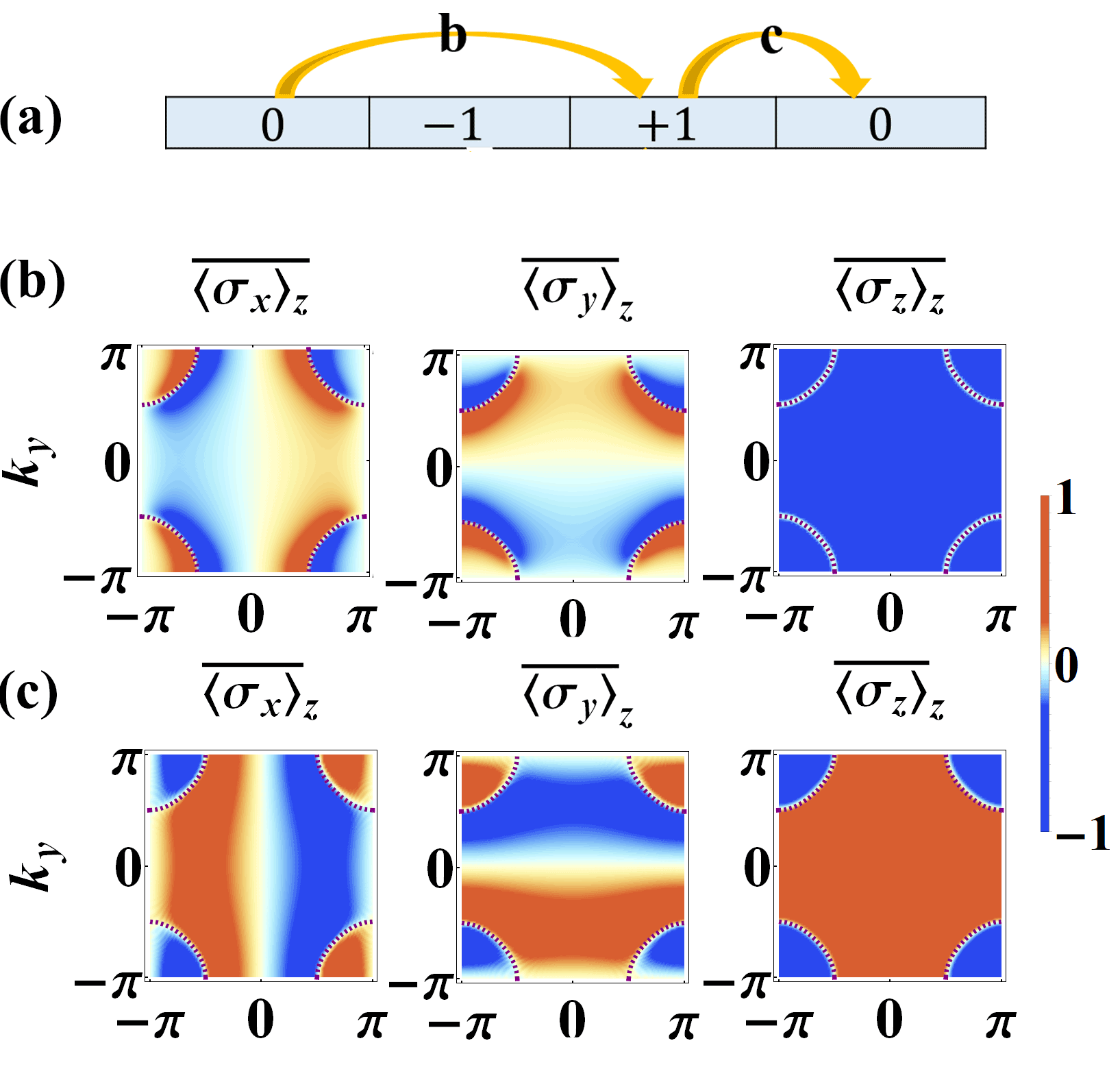, width=3.4in}
	\caption{(a) The specific quenching processes in (b) and (c). "0" represents the trivial phase, and other numbers represent the topological phases with different topological invariants. (b) The TASP after suddenly quenching $h_z$ axis from a trivial phase  to a topological phase with $C=+1$. The system is quenched from $m_z=20$ to $m_z=-1$ with $5t_{so}=t_0=1$. (c) The TASP after suddenly quenching $h_z$ axis from a topological phase  with $C=+1$ to a trivial phase. The system is quenched from $m_z=-1$ to $m_z=-10$ with $5t_{so}=t_0=1$.}
	\label{fig:tr_ts}
\end{figure}

For sudden quench, the quenching process from a topological phase to a trivial phase may not be distinguished from other types of processes. We make a comparison between the process from a trivial phase with a infinite $h_0^{int}$ to a topological phase with finite $h_0^{f}$ and the process from a topological phase with a finite $h_0^{int}$ to a trivial phase with a large $h_0^{f}$$(h_0^f>>h_1^2+h_2^2)$. For the former process, the TASP becomes $-\frac{{h^{f}_0}h_i^f}{{\ve_{f}}^2}$. Thus, only one common ring with $h_0^f=0$ (BIS) can be found in all the components of TASP. For the latter process,
the TASP becomes $-\frac{{h^{int}_0}h_i^f}{{\ve_{int}}{\ve_{f}}}$ and $h_0^f$ cannot be zero. Thus, only one common ring with ${h_0^{int}}=0$ can be found in all the components of TASP. As shown in Fig.~\ref{fig:tr_ts}(b) and Fig.~\ref{fig:tr_ts}(c), the rings in this two quenching processes have same features, and thus the two processes cannot be distinguished in sudden quench. 

However, in the slow quench, for the former process, the rings in TASP are BIS and SIS. On BIS,  $\overline{\la {{\sigma_{z}}}\ra}=0$ but $\overline{\la {{\sigma_{x,y}}}\ra}\ne0$. For the latter process, the ring in TASP is SIS, on which all the components of TASP $\overline{\la{{\sigma_{x,y,z}}}\ra}$ vanish. Thus, this two processes can be distinguished.

We conclude that, the difference between BIS and SIS always make the processes from a topological phase to a trivial phase distinguishable from other types of processes in the slow quench.  Compared with sudden quench, this is also the unique advantage of slow quench.

\section{Conclusions}\label{section:7i}
In summary, taking the 2D Chern insulator as an example, we uncover how the dynamical characterization is performed in different types of quenching processes. We show that the initial phase can be characterized under nonadiabatic quench dynamics. No matter what intermediate phase the quenching process undergoes, the TASP records the topological information of the initial phase and final phase.  Compared with the sudden quench, the processes from a topological phase to a trivial phase can always be distinguished from other type of processes. It is worthwhile that the dynamical characterization scheme here is only based on the TASP, and thus one can expect our characterization schemes may provide reference for future experiments.

\section*{Acknowledgements}
 This work was supported by the National Key Research and Development Program of Ministry of Science and Technology (No. 2021YFA1200700), National Natural Science Foundation of China (No. 11905054, and No.11804122), the China Postdoctoral Science Foundation (Grant No. 2021M690970)  and  the Fundamental Research Funds for the Central Universities from China. 

\begin{appendix}

	\section{The TASP under nonadiabatic quench dynamics}\label{appendix A}
{\it In sudden quench}, the Hamiltonian of two-level  system can be written as:	
\be
		{{\cal H}}={\bh}\cdot {\bm \sigma}=\left(\begin{array}{cc}
		 \varepsilon \cos\theta  &  \varepsilon \sin\theta e^{- i\varphi}   \\
		\varepsilon \sin\theta e^{i \varphi}      &  - \varepsilon \cos\theta
	\end{array}
	\right). \label{eq:LZ}
	\ee
Here, $\ve=\sqrt{{h_0}^2+{h_1}^2+{h_2}^2}$ is the energy of the Hamiltonian. The parameters $\theta$ and $\varphi$ are space angles that describe the direction of the effective field $\bh$.  In the following, we use superscripts/subscripts $``f"$ and $``int"$ to represent the parameters of the final Hamiltonian ${\cal H}_f$ and initial Hamiltonian  ${\cal H}_{int}$, respectively. 
  
The evolution of state vector can be written as: $|\psi(t) \ra =e^{-i{{\cal H}_{f}}t}|\psi(0) \ra$. Furthermore, an arbitrary initial state can be expressed as the superposition state of eigenstates of the final Hamiltonian: $|\psi(0) \ra =C_1|+ \ra+C_2|-\ra$. After some algebras, we show 
$\overline{\la {\sigma_i} \ra}=({C_1}^2-{C_2}^2)\frac{h_i^f}{\ve_f}$. Then if one choose the initial state as the ground state of the initial Hamiltonian, the TASP will become
\begin{small}
\be
\overline{\la {\sigma_i} \ra}=[-\cos(\theta_f-\theta_{int})-\sin{\theta_f}\sin{\theta_{int}}(\cos(\varphi_f-\varphi_{int})-1)]\frac{h_i^f}{\ve_f}.\nonumber	
\ee	
\end{small}
If one further choose $h_0$ as the quenching axis, $h_0$ will change from $h_0^{int}$ to $h_0^{f}$, and thus $\theta$ will change from $\theta_{int}$ to $\theta_{f}$. However, in both initial and final Hamiltonians, $h_1$ and $h_2$ are unchanged, and thus, $\varphi_{int}$= $\varphi_{f}$. At this time, the TASP is given in a simple form:
\be 
\begin{split}
	\overline{\la {\sigma_i} \ra}&=
	[-\cos(\theta_f-\theta_{int})]\frac{h_i^f}{\ve_f}\\
	&=-\frac{h_0^fh_0^{int}+h_1^2+h_2^2}{\ve_{f}\cdot\ve_{int}}\frac{h_i^f}{\ve_f}.
\end{split}	
\label{eq:sq}
\ee
This is a general result that can be applied to an arbitrary process.  

For the Chern insulator, $h_0$, $h_1$, and $h_2$ here equal to $h_z$, $h_x$, and $h_y$, respectively. When the system is quenched from $m_z^{int}$ to $m_z^f$, the TASP may be vanished at the regions with $h_z^fh_z^{int}+h_x^2+h_y^2=0$ and $h_z^f=0$, i.e., the SIS$_{g=0}$ and BIS. If SIS$_{g=0}$ has intersection points with the line $k=k_x=k_y$, the $t_{so}$ should satisfy the following equation:
\begin{small}
\be
2{t^2_{so}} {\sin}^2 k+m_z^{int}m_z^{f}-2(m_z^{int}+m_z^{f})\cos k+4{\cos}^2 k=0.\quad
\label{eq:SIS0}
\ee
\end{small}

For the case $m_z^{int}=-m_z^{f}=m_z$, we arrive at ${\sin}^2 k=\frac{m^2_z-4}{2t^2_{so}-4}$, and thus $-\frac{m_z}{\sqrt{2}}\leq t_{so}\leq \frac{m_z}{\sqrt{2}}$.  Under the critical value $\frac{m_z}{\sqrt{2}}$, SIS$_{g=0}$ has intersection points with the line $k=k_x=k_y$, and both the initial and final phase can be characterized. Beyond the critical value $\frac{m_z}{\sqrt{2}}$, SIS$_{g=0}$ has no intersection points with the line $k=k_x=k_y$, and both the initial and final phase cannot be characterized. For $m_z=1$, the critical value is  $\frac{\sqrt{2}}{2}\approx 0.7$.

{\it In slow quench}, a time-dependent Hamiltonian with quench protocol $g/t$ is given as:
\be
{{{\cal H}(t)}}={\bh(t)}\cdot {\bm \sigma}=\left(\begin{array}{cc}
	\frac{g}{t}+\varepsilon \cos\theta  &  \varepsilon \sin\theta e^{- i\varphi}   \\
	\varepsilon \sin\theta e^{i \varphi}      &  - \frac{g}{t}-\varepsilon \cos\theta
\end{array}
\right).
\ee
 
The evolution of the state vector is fully determined by the Schr\"odinger equation  $i\hbar\frac{d}{dt}|\psi(\bm k,t)\ra={{\cal{H}}(t)}|\psi(\bm k,t)\ra$ after preparing an initial state $|\psi(\bm k,t_{int})\ra$. In general, up to a total phase factor $\phi$, the state vector at the long-time limit can be given as \cite{2022us2,2020us1}: 
\be
|\psi(t) \ra = \sqrt{P_u}  e^{-i\ve_f t} |+ \ra +  \sqrt{P_d} e^{i\ve_f t + \phi} |{}-\ra.
\ee
Here,  $\phi=0$ in the case of sudden quench. $|\pm\ra$ satisfy the eigen equation $H_f|\pm\ra=\pm\ve_f|\pm\ra$.

Then the TASP can  be obtained by using the relation $\la \pm| H_f|\pm\ra=\pm\ve_f$ and ignoring the terms dependent on the time,
\be
\overline{\la {\sigma_i} \ra} =(P_u-P_d)\frac{\partial \ve_f}{\partial h_i^f}=(P_u-P_d)\frac{h_i^f}{\ve_f}.
\ee 

Thus, there exit some common spin polarizations with $(P_u-P_d)=0$ in all the components of TASP.  Here, we define the direction $k_{\perp}$ to be perpendicular to the $(P_u-P_d)=0$. Thus, the dynamical field on the $(P_u-P_d)=0$ can be shown to be proportional to the spin-orbit field $\bm {h_{so}}$:
\be
-\displaystyle_{\partial{k_{\perp}}}{\overline{\la \sigma_{so,i}\ra}}\propto\lim_{{k_{\perp}} \rightarrow 0}\frac{1}{2k_{\perp}}\frac{h_{so,i}+O(k_{\perp})}{{\ve_f}+O(k_{\perp})}2k_{\perp}=\frac{h_{so,i}}{{\ve_f}}.\quad
\label{eq:A1}
\ee
with $i=1,2$. $\bm {h_{so}}=(h_1,h_2)$ is a two components vector. Similarly, one can also show the dynamical field is proportional to spin-orbit field in sudden quench. 
\section{The  charge configurations in different processes and the translation of the TASP between two reversed processes}\label{appendix B}

\begin{figure}[htbp] 
	\centering
	\epsfig{file=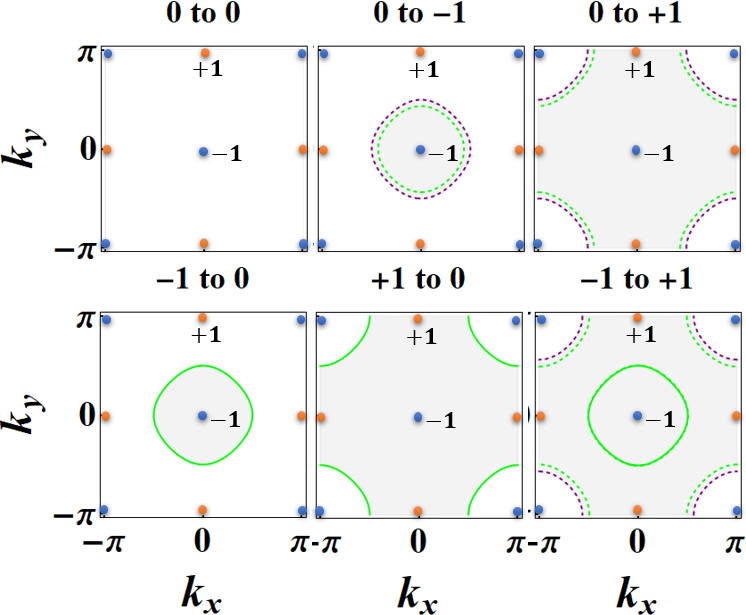, width=3.4in}
	\caption{ The schematic diagram of topological charges. The gray region represents the region enclosed by 	the rings. Here, $t_{so}=0.2t_0.$}
	\label{fig:topological_charge}
\end{figure}

{\it The  charge configurations.} In general, the vector field can be decomposed into $h_0=h_z$ and $\bm {h}_{\bm {so}}=(h_y,h_x)$. $\bm {h}_{\bm {so}}=0$ are the positions of topological charges. As shown in Fig.~\ref{fig:topological_charge}, we plot the different charge configurations in all types of quenching processes. When we quench the system from a trivial phase to another trivial phase, there is no rings, and thus no topological charge is enclosed by the rings. When we quench the system from a trivial phase to a topological phase with $C=-1$, a topological charge $-1$ located at $k_x=k_y=0$ is enclosed by both BIS and SIS, and thus gives the Chern number $C=-1$.
Similarly, another two topological charges $+1$ at $(0,\pi)$ and $(\pi,0)$ are also included when the system is quenched from a trivial phase to a topological phase with $C= +1$, which renders the Chern number being changed to $1+1-1=+1$. Furthermore, when we quench the system from a topological phase to a trivial phase, a topological charge $-1(+1)$is enclosed by the SIS, and thus gives the Chern number $C=-1(+1)$. Finally, when we quench the system from a topological phase to another topological phase, a topological charge $-1$ is enclosed by the ISIS and another topological charge $+1$ is enclosed by both BIS and FSIS. Therefore, both the initial and final phase of the system can be characterized. These results are consistent with those given in the main text by the dynamical field.

\begin{figure}[htbp] 
	\centering
	\epsfig{file=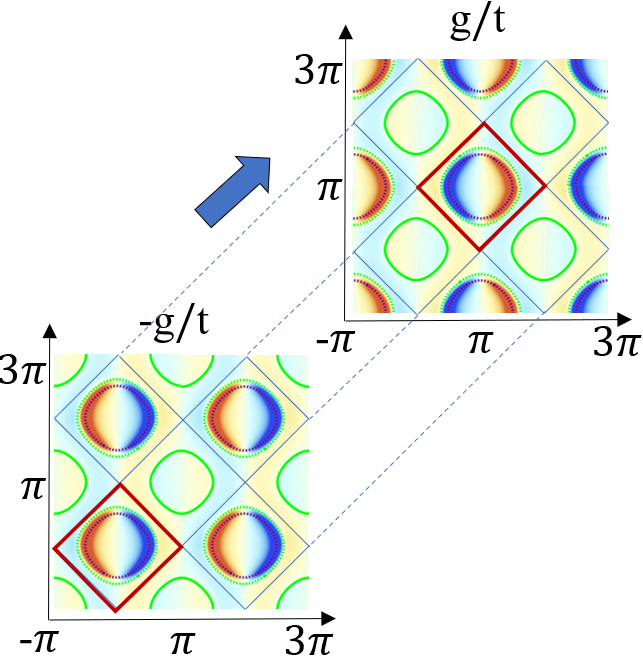, width=3.4in}
	\caption{(a) The schematic diagram of translation between two reversed processes. The arrow represents the direction of translation.}
	\label{fig:2D}
\end{figure}

{\it The translation of the TASP.} As shown in Fig.~\ref{fig:2D}, we show a schematic diagram of translation between two reversed processes. One can easily see that the TASP between protocols $g/t$ and $-g/t$ in the square region marked in red have the relation $
\overline{\la { {\sigma}_i(\bm k)}\ra}_{g/t}= -\overline{\la { {\sigma}_i(\bm k+\pi)}\ra}_{{-g/t}}$. 
\section{The  TASP of topological insulator with high integer invariant}\label{appendix C}
\begin{figure}[htbp] 
	\centering
	\epsfig{file=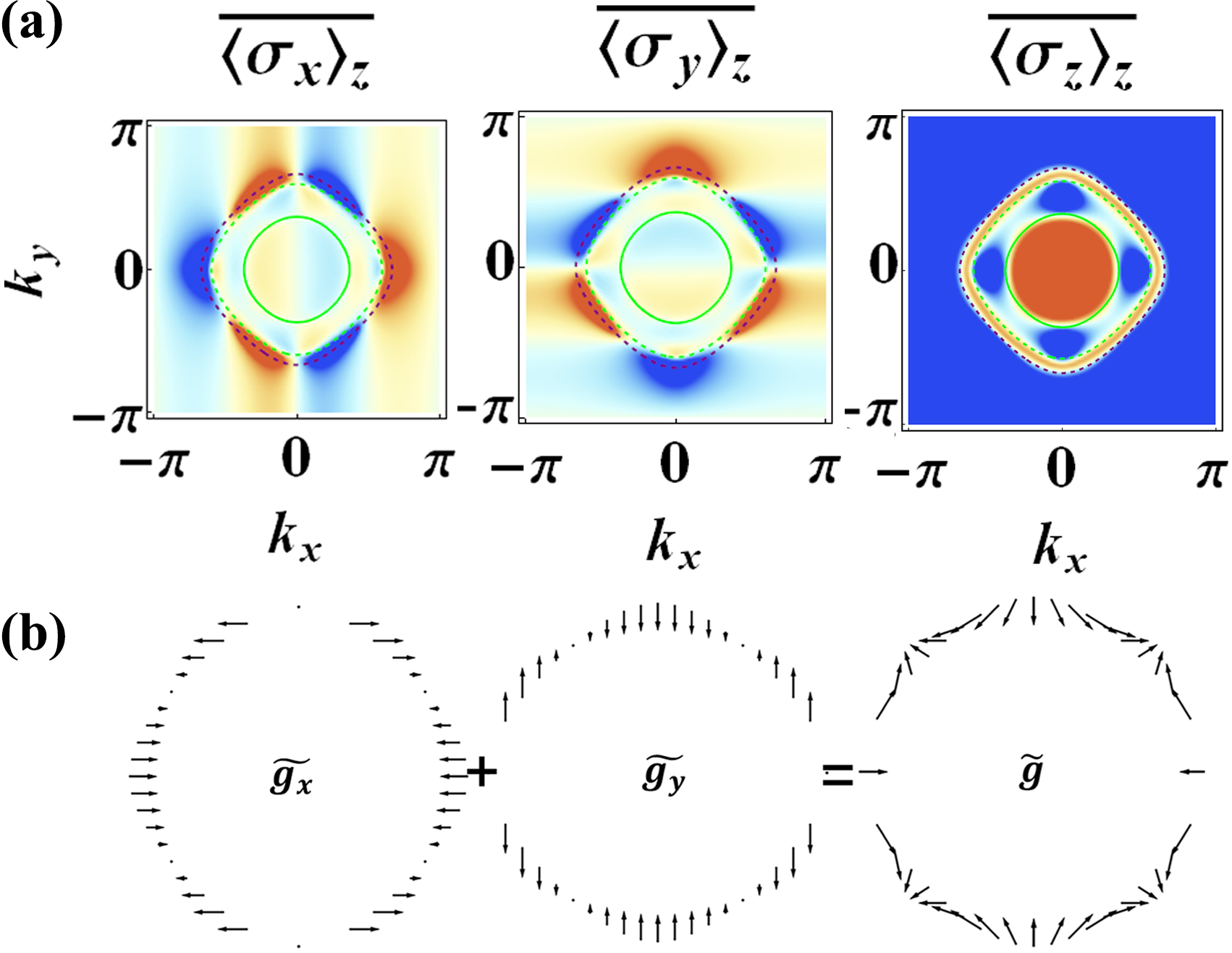, width=3.4in}
	\caption{(a) The  TASP of topological insulator with high integer invariant. (b) The dynamical field on the BIS and FSIS. The results are obtained by a slow quench from $m_z= 1.5{t_0}$ to $0.5t_0$ with $t_{so}=0.2t_0$. }
	\label{fig:high}
\end{figure}

We consider a model with the effective field $\bm{h}(\bm k)=(t_{so}\sin 2k_x, t_{so}\sin 2k_y, m_z-  t_{0}\cos k_x-t_{0} \cos k_y)$. While the trivial phase corresponds to $|m_z|>2t_0$, the topological phases are distinguished as: (i) $t_0<m_z<2t_0$ with the Chern number $C=-1$; (ii) $0<m_z<t_0$ with $C=3$; (iii) $-t_0<m_z<0$ with $C=-3$; (iv) $-2t_0<m_z<-t_0$ with $C=1$.  We provide the results of quenching the system from the phase with $C=-1$ to the phase with $C=3$ [see Fig.~\ref{fig:high}].  Three rings also appear in the TASP, but the corresponding dynamical field on BIS and FSIS winds three times, implying the final topological invariant $C=3$. The dynamical field on the ISIS gives an initial topological invariant $-1$. Thus, both the initial and final phases can be characterized.
\end{appendix}

\end{document}